\documentclass[american,english,prl, twocolumn]{revtex4}
\usepackage[T1]{fontenc}
\usepackage[latin9]{inputenc}
\usepackage{color}
\usepackage{babel}
\usepackage{amstext}
\usepackage{amssymb}
\usepackage{stmaryrd}
\usepackage{esint}
\usepackage[pdfusetitle,
 bookmarks=true,bookmarksnumbered=false,bookmarksopen=false,
 breaklinks=false,pdfborder={0 0 1},backref=false,colorlinks=false]
 {hyperref}

\makeatletter
\@ifundefined{textcolor}{}
{%
 \definecolor{BLACK}{gray}{0}
 \definecolor{WHITE}{gray}{1}
 \definecolor{RED}{rgb}{1,0,0}
 \definecolor{GREEN}{rgb}{0,1,0}
 \definecolor{BLUE}{rgb}{0,0,1}
 \definecolor{CYAN}{cmyk}{1,0,0,0}
 \definecolor{MAGENTA}{cmyk}{0,1,0,0}
 \definecolor{YELLOW}{cmyk}{0,0,1,0}
}

\usepackage{tikz}

\makeatother

\begin{document}
\title{Computation of the semiclassical outflux emerging from a collapsing
spherical null shell}
\author{Amos Ori}
\email{amos@physics.technion.ac.il}

\affiliation{Department of Physics, Technion, Haifa 32000, Israel}
\author{Noa Zilberman}
\email{nz3745@princeton.edu}

\affiliation{Princeton Gravity Initiative, Princeton University, Princeton, New
Jersey 08544, USA}
\date{\today}
\begin{abstract}
\noindent We consider a minimally coupled, massless quantum scalar
field $\hat{\Phi}$ propagating in the background geometry of a four-dimensional
black hole formed by the collapse of a spherical thin null shell,
with a Minkowski interior and a Schwarzschild exterior. The field
is taken in the natural ``in'' vacuum state, namely, the quantum
state in which no excitations arrive from past null infinity. Within
the semiclassical framework, we analyze the vacuum polarization $\left\langle \hat{\Phi}^{2}\right\rangle _{\text{ren}}$
and the energy outflux density $\left\langle \hat{T}_{uu}\right\rangle _{\text{ren}}$
(where $u$ is the standard null Eddington coordinate) just outside
the shell. Using the point-splitting method, we derive closed-form
analytical expressions for both of these semiclassical quantities.
In particular, our result for $\left\langle \hat{T}_{uu}\right\rangle _{\text{ren}}$
reveals that it vanishes like $(1-2M_{0}/r_{0})^{2}$ as the shell
collapses toward the event horizon, where $M_{0}$ is the shell's
mass and $r_{0}$ is the value of the area coordinate $r$ at the
evaluation point. This confirms that, along a late-time outgoing null
geodesic (i.e., one that emerges from the shell very close to the
event horizon and propagates toward future null infinity), the outflux
gradually evolves (from virtually zero) up to its final Hawking-radiation
value while the geodesic traverses the strong-field region (rather
than the Hawking-radiation outflux being emitted entirely from the
collapsing shell, which would lead to significant backreaction effects).
\end{abstract}
\maketitle

\paragraph*{Introduction.}

When a compact object collapses to form a black hole (BH), quantum
field theory predicts the emission of massless particles (photons
and gravitons), which carry energy to future null infinity (FNI).
The emission process relaxes, at a later time, to a steady state known
as \emph{Hawking radiation} \citep{Hawking:1974,Hawking:1975}.

Where does this Hawking-radiation energy come from? This has been
a long-standing question \citep{Unruh:1977,ChenUnruhWuYeom:2018}.
It is quite obvious that it emerges out of the strong-field region,
but it is less clear exactly from what part of that region it originates.
A widely accepted view (see, e.g., Ref. \citep{Unruh:1977,ChenUnruhWuYeom:2018}),
to which we shall refer as \emph{option }(\emph{i}), is that the outgoing
radiation starts from zero at the event horizon (EH) and gradually
develops over a fairly broad strong-field region, until the area coordinate
$r$ reaches a few times the BH mass $M$. However, there has also
been an alternative viewpoint (\emph{option }(\emph{ii})) suggesting
that this energy outflux actually emerges directly from the collapsing
body itself, and propagates in a conserved (or approximately conserved)
manner through the surrounding strong-field vacuum region until it
arrives FNI as Hawking radiation (see e.g. \citep{Davies:1976,Gerlach:1976,Boulware:1976,KawaiMatsuoYokokura:2013,Ho:2015,KawaiYokokura:2016,Ho:2016,MaskalaniecSikorski:2024}).

For simplicity, let us consider a spherically-symmetric collapsing
object, surrounded by a vacuum region whose geometry is described
by the Schwarzschild metric $ds^{2}=-f\,dt^{2}+f^{-1}dr^{2}+r^{2}d\Omega^{2}$,
where $f\equiv1-2M/r$ and $d\Omega^{2}\equiv d\theta^{2}+\sin^{2}\theta\,d\varphi^{2}$
\footnote{We use general-relativistic units $c=G=1$.}. We shall focus
here on the domain $r>2M$ of this Schwarzschild geometry. We define
the two null Eddington coordinates $u,v$ by $v\equiv t+r_{*}$ and
$u\equiv t-r_{*}$, where $r_{*}\equiv r+2M\ln\left(r/2M-1\right)$
(in Minkowski, $r_{*}=r$). The energy outflux density is then represented
by $\left\langle \hat{T}_{uu}\right\rangle _{\text{ren}}$, that is,
the outgoing null component of the \emph{renormalized stress-energy
tensor} (RSET) $\left\langle \hat{T}_{\mu\nu}\right\rangle _{\text{ren}}(x)$.

Let us focus in this qualitative discussion on the ``late-time domain''
(namely, the domain of sufficiently large $u$), where the quantum
outflux (at FNI) has already settled down to its constant value. In
that domain,
\[
\lim_{r\to\infty}4\pi r^{2}\left\langle \hat{T}_{uu}\right\rangle _{\text{ren}}=\text{const}\equiv F_{H}\,.
\]
The \emph{Hawking outflux} $F_{H}$ takes the general form $F_{H}=\hbar P/M^{2}$,
where $P$ is a dimensionless constant (involving the so-called \emph{greybody
factor}). According to option (\emph{ii}) above \textendash{} in which
the outflux
\[
F\left(u,v\right)\equiv4\pi r^{2}\left\langle \hat{T}_{uu}\right\rangle _{\text{ren}}(u,v)
\]
is essentially conserved while propagating along an outgoing null
geodesic \textendash{} $F$ is constant at large $u$ (either exactly
or approximately) not only at FNI, but also everywhere in the vacuum
region surrounding the collapsing object. In that case we have there
\[
\left\langle \hat{T}_{uu}\right\rangle _{\text{ren}}^{(ii)}(x)=F_{H}/4\pi r^{2}\,,\,\,\,\,\,\,\,\,(u\gg M)\,.
\]
The index ``$(ii)$'' was added here to relate this expression specifically
to option (\emph{ii}) (which we will argue is not the correct semiclassical
scenario). While this result might seem appealing due to its simplicity,
it actually entails drastic consequences, as we now briefly discuss.

At sufficiently late $u$, the surface of the collapsing object has
already approached its near-EH $r$ value, $r\approx2M$. We may therefore
write, for the region immediately outside the object's surface,
\[
\left\langle \hat{T}_{uu}\right\rangle _{\text{ren}}^{(ii)(\text{surface})}\approx\text{const}\approx F_{H}/4\pi\left(2M\right)^{2}\,,\,\,\,\,(u\gg M)\,.
\]
Recall, however, that the coordinate $u$ becomes singular on approaching
the EH (where $u\to\infty$). To properly assess the physical content
of this outflux-density expression, we transform it to the regular,
Kruskal coordinate $U\equiv-M\exp(-u/4M)$. The corresponding regularized
outflux-density expression becomes
\[
\left\langle \hat{T}_{UU}\right\rangle _{\text{ren}}^{(ii)(\text{surface})}=\left(F_{H}/\pi M^{2}\right)\exp(u/2M)\,,\,\,\,\,(u\gg M)\,.
\]
Evidently, this quantity \emph{diverges} on approaching the EH, which
would potentially lead to drastic backreaction effects \textendash{}
such as entirely halting the collapse process \citep{Gerlach:1976,Boulware:1976,KawaiMatsuoYokokura:2013,Mersini:2014(HH),Mersini:2014(Unruh),Ho:2015,KawaiYokokura:2016,Ho:2016,MaskalaniecSikorski:2024}
and/or preventing the EH formation (for various scenarios of ``horizon
avoidance'' in semiclassical collapse see, e.g., \citep{Gerlach:1976,KawaiMatsuoYokokura:2013,Ho:2015,KawaiYokokura:2016,Ho:2016,MaskalaniecSikorski:2024,Mersini:2014(HH),Mersini:2014(Unruh),BMT:2016,BMT:2017a,BMT:2017b}).

In the absence of concrete RSET calculations (in 4D gravitational
collapse), it is hard to prove which of the above two options is the
correct one. Our main goal in this letter is to present such a concrete
RSET analysis.

We hereafter model the collapse process by a thin null massive shell,
with an empty, flat interior (more details below). For simplicity
we consider a minimally-coupled massless scalar field $\Phi(x)$.
Several authors (see, e.g., Refs. \citep{AndersonSiahmazginClarkFabbri:2020,SiahmazgiAndersonClarkFabbri:2021,SiahmazgiAndersonFabbri:2025}
and references within) have already elaborated on this model, but
so far, to our knowledge, a concrete result for $\left\langle \hat{T}_{uu}\right\rangle _{\text{ren}}$
in the strong-field region has not been obtained. We shall derive
a simple, explicit, analytical expression for $\left\langle \hat{T}_{uu}\right\rangle _{\text{ren}}$
in the Schwarzschild region just outside the collapsing shell. Our
result (presented in Eq. (\ref{eq: result_Tuu})) demonstrates that
option (\emph{i}) above is the correct one.

\paragraph{The collapsing-shell model. \label{sec:  Collapsing-shell-model}}

\begin{figure}
%% LyX 2.4.2.1 created this file.  For more info, see https://www.lyx.org/.
%% Do not edit unless you really know what you are doing.

\tikzset{every picture/.style={line width=0.75pt}} %set default line width to 0.75pt        

\scalebox{0.8}{

\begin{tikzpicture}[x=0.75pt,y=0.75pt,yscale=-1,xscale=1]
%uncomment if require: \path (0,244); %set diagram left start at 0, and has height of 244

%Shape: Wave [id:dp9981609703612297] 
\draw   (122.03,10) .. controls (122.68,11.37) and (123.31,12.67) .. (124.03,12.67) .. controls (124.75,12.67) and (125.38,11.37) .. (126.03,10) .. controls (126.68,8.63) and (127.31,7.33) .. (128.03,7.33) .. controls (128.75,7.33) and (129.38,8.63) .. (130.03,10) .. controls (130.68,11.37) and (131.31,12.67) .. (132.03,12.67) .. controls (132.75,12.67) and (133.38,11.37) .. (134.03,10) .. controls (134.68,8.63) and (135.31,7.33) .. (136.03,7.33) .. controls (136.75,7.33) and (137.38,8.63) .. (138.03,10) .. controls (138.68,11.37) and (139.31,12.67) .. (140.03,12.67) .. controls (140.75,12.67) and (141.38,11.37) .. (142.03,10) .. controls (142.68,8.63) and (143.31,7.33) .. (144.03,7.33) .. controls (144.75,7.33) and (145.38,8.63) .. (146.03,10) .. controls (146.68,11.37) and (147.31,12.67) .. (148.03,12.67) .. controls (148.75,12.67) and (149.38,11.37) .. (150.03,10) .. controls (150.68,8.63) and (151.31,7.33) .. (152.03,7.33) .. controls (152.75,7.33) and (153.38,8.63) .. (154.03,10) .. controls (154.68,11.37) and (155.31,12.67) .. (156.03,12.67) .. controls (156.75,12.67) and (157.38,11.37) .. (158.03,10) .. controls (158.68,8.63) and (159.31,7.33) .. (160.03,7.33) .. controls (160.75,7.33) and (161.38,8.63) .. (162.03,10) .. controls (162.68,11.37) and (163.31,12.67) .. (164.03,12.67) .. controls (164.75,12.67) and (165.38,11.37) .. (166.03,10) .. controls (166.68,8.63) and (167.31,7.33) .. (168.03,7.33) .. controls (168.75,7.33) and (169.38,8.63) .. (170.03,10) .. controls (170.68,11.37) and (171.31,12.67) .. (172.03,12.67) .. controls (172.75,12.67) and (173.38,11.37) .. (174.03,10) .. controls (174.68,8.63) and (175.31,7.33) .. (176.03,7.33) .. controls (176.75,7.33) and (177.38,8.63) .. (178.03,10) .. controls (178.68,11.37) and (179.31,12.67) .. (180.03,12.67) .. controls (180.75,12.67) and (181.38,11.37) .. (182.03,10) .. controls (182.68,8.63) and (183.31,7.33) .. (184.03,7.33) .. controls (184.75,7.33) and (185.38,8.63) .. (186.03,10) .. controls (186.68,11.37) and (187.31,12.67) .. (188.03,12.67) .. controls (188.75,12.67) and (189.38,11.37) .. (190.03,10) .. controls (190.68,8.63) and (191.31,7.33) .. (192.03,7.33) .. controls (192.75,7.33) and (193.38,8.63) .. (194.03,10) .. controls (194.68,11.37) and (195.31,12.67) .. (196.03,12.67) .. controls (196.75,12.67) and (197.38,11.37) .. (198.03,10) .. controls (198.68,8.63) and (199.31,7.33) .. (200.03,7.33) .. controls (200.75,7.33) and (201.38,8.63) .. (202.03,10) .. controls (202.68,11.37) and (203.31,12.67) .. (204.03,12.67) .. controls (204.75,12.67) and (205.38,11.37) .. (206.03,10) .. controls (206.68,8.63) and (207.31,7.33) .. (208.03,7.33) .. controls (208.75,7.33) and (209.38,8.63) .. (210.03,10) .. controls (210.68,11.37) and (211.31,12.67) .. (212.03,12.67) .. controls (212.75,12.67) and (213.38,11.37) .. (214.03,10) .. controls (214.68,8.63) and (215.31,7.33) .. (216.03,7.33) .. controls (216.75,7.33) and (217.38,8.63) .. (218.03,10) .. controls (218.68,11.37) and (219.31,12.67) .. (220.03,12.67) .. controls (220.75,12.67) and (221.38,11.37) .. (222.03,10) .. controls (222.68,8.63) and (223.31,7.33) .. (224.03,7.33) .. controls (224.75,7.33) and (225.38,8.63) .. (226.03,10) .. controls (226.68,11.37) and (227.31,12.67) .. (228.03,12.67) .. controls (228.75,12.67) and (229.38,11.37) .. (230.03,10) .. controls (230.68,8.63) and (231.31,7.33) .. (232.03,7.33) .. controls (232.75,7.33) and (233.38,8.63) .. (234.03,10) .. controls (234.68,11.37) and (235.31,12.67) .. (236.03,12.67) .. controls (236.75,12.67) and (237.38,11.37) .. (238.03,10) .. controls (238.13,9.79) and (238.23,9.58) .. (238.33,9.37) ;
%Shape: Boxed Line [id:dp6895446615047802] 
\draw [color={rgb, 255:red, 74; green, 144; blue, 226 }  ,draw opacity=1 ]   (127.71,225.73) -- (272.75,80.73) ;
\draw [shift={(274.17,79.32)}, rotate = 135.01] [color={rgb, 255:red, 74; green, 144; blue, 226 }  ,draw opacity=1 ][line width=0.75]    (10.93,-3.29) .. controls (6.95,-1.4) and (3.31,-0.3) .. (0,0) .. controls (3.31,0.3) and (6.95,1.4) .. (10.93,3.29)   ;
%Straight Lines [id:da1853837077854701] 
\draw [color={rgb, 255:red, 74; green, 144; blue, 226 }  ,draw opacity=1 ]   (235,117.34) -- (199.38,79.79) ;
\draw [shift={(198,78.34)}, rotate = 46.51] [color={rgb, 255:red, 74; green, 144; blue, 226 }  ,draw opacity=1 ][line width=0.75]    (10.93,-3.29) .. controls (6.95,-1.4) and (3.31,-0.3) .. (0,0) .. controls (3.31,0.3) and (6.95,1.4) .. (10.93,3.29)   ;
%Straight Lines [id:da9818538558300707] 
\draw    (121.95,241.33) -- (294.67,67.67) -- (238.67,9.19) -- (121.95,124.52) ;
%Straight Lines [id:da6958818607878385] 
\draw [line width=0.75]    (121.95,7.71) -- (121.95,241.33) ;
%Straight Lines [id:da29889475741178284] 
\draw [color={rgb, 255:red, 208; green, 2; blue, 27 }  ,draw opacity=1 ][line width=1.5]    (121.95,8.19) -- (235.33,127.14) ;

% Text Node
\draw (188.52,84.82) node [anchor=north west][inner sep=0.75pt]  [font=\normalsize,color={rgb, 255:red, 208; green, 2; blue, 27 }  ,opacity=1 ,rotate=-45]  {${\textstyle v=v_{0}}$};
% Text Node
\draw (196,61) node [anchor=north west][inner sep=0.75pt]  [font=\large,color={rgb, 255:red, 74; green, 144; blue, 226 }  ,opacity=1 ,rotate=-45]  {$\tilde{u}$};
% Text Node
\draw (270,72) node [anchor=north west][inner sep=0.75pt]  [font=\large,color={rgb, 255:red, 74; green, 144; blue, 226 }  ,opacity=1 ,rotate=-315]  {$v$};
% Text Node
\draw (118.67,119.14) node [anchor=north west][inner sep=0.75pt]  [font=\normalsize,rotate=-90]  {$( r=0)$};
% Text Node
\draw (286.91,7.3) node [anchor=north west][inner sep=0.75pt]  [font=\normalsize,rotate=-45] [align=left] {FNI};
% Text Node
\draw (204.18,166.31) node [anchor=north west][inner sep=0.75pt]  [font=\normalsize,rotate=-315] [align=left] {PNI};
% Text Node
\draw (160.92,14.92) node [anchor=north west][inner sep=0.75pt]  [font=\normalsize]  {$r=0$};
% Text Node
\draw (264.86,6.35) node [anchor=north west][inner sep=0.75pt]  [font=\normalsize,rotate=-45]  {${\textstyle \left( r\rightarrow \infty \right)}$};
% Text Node
\draw (227.72,138.78) node [anchor=north west][inner sep=0.75pt]  [font=\normalsize,rotate=-315]  {${\textstyle \left( r\rightarrow \infty \right)}$};
% Text Node
\draw (118.67,70.71) node [anchor=north west][inner sep=0.75pt]  [font=\normalsize,rotate=-90] [align=left] {center};
% Text Node
\draw (203.31,50.51) node [anchor=north west][inner sep=0.75pt]  [font=\normalsize,rotate=-315] [align=left] {EH};

\end{tikzpicture}

}

\caption{The Penrose diagram of a spherical collapsing null shell spacetime.
The red line represents the null shell. Above (below) the red line
lies the shell's exterior (interior), described by the Schwarzschild
(Minkowski) geometry. Blue arrows indicate the global $\tilde{u}$
and $v$ coordinates (outside the EH). \label{fig:The-Penrose-diagram}}
\end{figure}
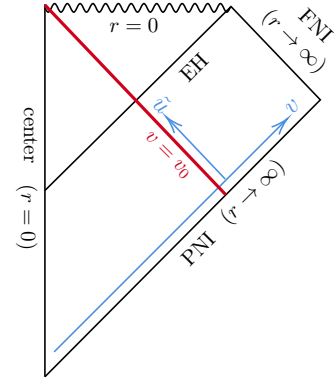

We consider a spherically-symmetric collapsing thin null shell, of
mass $M_{0}$, propagating inward along an incoming null ray $v=v_{0}$
(see Fig. \ref{fig:The-Penrose-diagram}).

The geometry outside the shell (namely at $v>v_{0}$) is Schwarzschild
with $M=M_{0}$, while inside the shell ($v<v_{0}$) it is Minkowski
($M=0$), that is, $M=M_{0}\Theta(v-v_{0})$, where $\Theta$ denotes
the standard step function. These two different geometries are continuously
matched across the shell \textendash{} that is, the overall metric
should be continuous (in appropriate coordinates).

The Schwarzschild geometry is given in double-null Eddington coordinates
$(u,v,\theta,\varphi)$ by $ds^{2}=-f\,du\,dv+r^{2}d\Omega^{2}$,
where $r$ is now viewed as a function of the coordinates $u,v$ (defined
implicitly via the relation $r_{*}(r)$ given above, as $r_{*}=\left(v-u\right)/2$).

As already mentioned, the metric across the thin shell is in principle
continuous (provided that one uses appropriate coordinates \textendash{}
which are themselves continuous across the shell, as we do below).
In particular, $r$ is necessarily continuous at the shell, and so
are the coordinates $v,\theta,\varphi$. Note, however, that if the
Eddington coordinate $u$ is used globally (i.e. at both sides), the
resultant metric is discontinuous. In particular, $-2g_{uv}$ is equal
to $f=1-(2M_{0}/r)\Theta(v-v_{0})$, wherein the step function $\Theta(v-v_{0})$
indicates discontinuity at the shell. To achieve metric continuity,
we introduce a new global null coordinate $\tilde{u}$ (defined outside
the EH) as follows: in the Schwarzschild region outside the shell
we set $\tilde{u}\equiv u$; and inside the shell we define $\tilde{u}\left(u\right)$
by simply requiring it to continuously match the aforementioned Schwarzschild's
$u$$(=\tilde{u})$. Explicitly, this continuous $\tilde{u}$ coordinate
turns out to be (see \citep{MassarParentani:1996,FabbriNavarroSalas:Book,AndersonSiahmazginClarkFabbri:2020})
\begin{equation}
\tilde{u}(u)\equiv u-4M_{0}\ln\left(\frac{v_{0}-u}{4M_{0}}-1\right)\Theta(v_{0}-v)\,.\label{eq: u_tilde-2}
\end{equation}
The metric is then continuous across the shell, with $g_{\tilde{u}v}$
being $-(1/2)(1-2M_{0}/r)$ at the shell (as approached from both
sides). For later use, we also note that inside the shell
\begin{equation}
\frac{du}{d\tilde{u}}=1-\frac{4M_{0}}{v_{0}-u}=1-\frac{2M_{0}}{r(u,v_{0})}\,,\,\,\,\,\,\,\,\,\,\,\,\,(v<v_{0})\,.\label{eq: du_tilde}
\end{equation}

\paragraph{Field modes and quantum state.}

Our real scalar field $\Phi$ satisfies the field equation $\Box\Phi=0$.
We decompose it into normalized mode solutions $\Phi_{\omega lm}(x)$,
defined by their asymptotic initial data at past null infinity (PNI):
\begin{equation}
\Phi_{\omega lm}(x)=\frac{1}{r\sqrt{4\pi\omega}}Y_{lm}(\theta,\varphi)\exp(-i\omega v)\,,\,\,\,\,(\text{PNI})\label{eq: modes}
\end{equation}
along with the usual regularity condition at $r=0$. Here $Y_{lm}(\theta,\varphi)$
are the spherical harmonics.

The \emph{quantum} field may then be expressed as
\begin{equation}
\hat{\Phi}(x)=\!\int_{0}^{\infty}\!d\omega\sum_{lm}\left[\Phi_{\omega lm}(x)\hat{a}_{\omega lm}+\Phi_{\omega lm}^{*}(x)\hat{a}_{\omega lm}^{\dag}\right],\label{eq:phi_qua}
\end{equation}
where $\hat{a}_{\omega lm}$ and $\hat{a}_{\omega lm}^{\dagger}$
are the annihilation and creation operators for the $\omega lm$ mode,
respectively. Our quantum state, denoted $|0\rangle$, is taken to
be the ``in'' vacuum state, namely the vacuum state with respect
to the mentioned positive-frequency modes $\Phi_{\omega lm}(x)$,
obeying $\hat{a}_{\omega lm}|0\rangle=0$ (indicating no excitations
originating from PNI).

The fact that $\Box\Phi_{\omega lm}=0$, combined with the metric
continuity, implies that the modes $\Phi_{\omega lm}(x)$ are continuous
across the shell.

In an entirely Minkowski spacetime (no collapsing shell), both $\left\langle \hat{\Phi}^{2}\right\rangle _{\text{ren}}$
and the RSET trivially vanish everywhere in the vacuum state $|0\rangle$.
In our model, spacetime remains Minkowski until the shell emerges
from PNI at $v=v_{0}$. By causality, this shell cannot affect physical
quantities at earlier $v$ values. Therefore, both $\left\langle \hat{\Phi}^{2}\right\rangle _{\text{ren}}$
and the RSET vanish throughout the shell's interior, namely at $v<v_{0}$.
However, at the shell's exterior, these quantities generally do not
vanish due to the existing curvature. Our goal in this paper is to
compute $\left\langle \hat{\Phi}^{2}\right\rangle _{\text{ren}}$
and $\left\langle \hat{T}_{uu}\right\rangle _{\text{ren}}$ at the
shell's Schwarzschild side, that is, at $v\to v_{0}^{+}$.

\paragraph{Computation of $\left\langle \hat{\Phi}^{2}\right\rangle _{\text{ren}}$.}

To renormalize $\left\langle \hat{\Phi}^{2}\right\rangle $, at a
generic point $x$ in Schwarzschild (or Minkowski), we use the point-splitting
procedure:
\begin{equation}
\left\langle \hat{\Phi}^{2}\right\rangle _{\text{ren}}(x)=\lim_{x'\to x}\left[G\left(x,x'\right)-G_{DS}(x,x')\right]\,,\label{eq: PointSplit}
\end{equation}
where $G(x,x')\equiv\left\langle \hat{\Phi}\left(x\right)\hat{\Phi}\left(x'\right)\right\rangle $
and $G_{DS}(x,x')$ is the counterterm \citep{DeWittBook:1965}.

In view of the spherical symmetry, we choose $\theta$-splitting \citep{LeviOriTheta:2016}.
Without loss of generality, we pick our point $x$ to be at $(u,v,\theta=0)$
and $x'$ at $(u,v,\theta=\varepsilon)$; hence $\left\langle \hat{\Phi}^{2}\right\rangle _{\text{ren}}(x)$
effectively reduces to $\left\langle \hat{\Phi}^{2}\right\rangle _{\text{ren}}(u,v)$,
$G(x,x')$ to $G(u,v;\varepsilon)$, and $G_{DS}(x,x')$ to $G_{DS}(r;\varepsilon)$,
\footnote{In our problem the quantum state is spherically symmetric but not
static (due to the collapse). However, the counterterms are independent
of the quantum state and therefore respect the local staticity of
the Schwarzschild metric, depending only on $r$ (and $\varepsilon$).} where $r=r(u,v)$ is to be understood. Equation (\ref{eq: PointSplit})
then reduces to
\begin{equation}
\left\langle \hat{\Phi}^{2}\right\rangle _{\text{ren}}(u,v)=\lim_{\varepsilon\to0}\left[G(u,v;\varepsilon)-G_{DS}(r;\varepsilon)\right]\,.\label{eq: theta_split}
\end{equation}
This equation applies everywhere in the Schwarzschild region at $v>v_{0}$,
and likewise throughout the Minkowski region at $v<v_{0}$.

We now choose a point on the shell, denoted $x_{\text{sh}}$ (at $\theta=0$),
with a certain $r$ value denoted $r_{0}$ and a $\tilde{u}$ value
(determined by $v_{0}$ and $r_{0}$) denoted $\tilde{u}_{0}$. We
wish to compute $\left\langle \hat{\Phi}^{2}\right\rangle _{\text{ren}}(x_{\text{sh}})$
when the point $x_{\text{sh}}$ is approached ``from the Schwarzschild
side of the shell''. We hereafter use a superscript ``$+$'' to
denote quantities evaluated in the Schwarzschild geometry outside
the shell, approaching $x_{\text{sh}}$ (i.e., with $v\to v_{0}^{+}$).
Applying Eq. (\ref{eq: theta_split}) for this purpose, we obtain
\footnote{This transition implicitly involves interchanging the $v\to v_{0}^{+}$
and $\varepsilon\to0$ limits, which we are unable to rigorously justify
(see Discussion).\label{fn:transition}}
\begin{equation}
\left\langle \hat{\Phi}^{2}\right\rangle _{\text{ren}}^{+}(x_{\text{sh}})=\lim_{\varepsilon\to0}\left[G^{+}(\tilde{u}_{0},v_{0};\varepsilon)-G_{DS}^{+}(r_{0};\varepsilon)\right].\label{eq: theta_split-Sch}
\end{equation}

We can now repeat the above procedure approaching from the Minkowski
side:
\begin{equation}
\left\langle \hat{\Phi}^{2}\right\rangle _{\text{ren}}^{-}(x_{\text{sh}})=\lim_{\varepsilon\to0}\left[G^{-}(\tilde{u}_{0},v_{0};\varepsilon)-G_{DS}^{-}(r_{0};\varepsilon)\right]\,,\label{eq: theta_split-Mink}
\end{equation}
where the superscript ``$-$'' refers to quantities evaluated at
the Minkowski side of the shell, approaching $x_{\text{sh}}$ (with
$v\to v_{0}^{-}$). Note, however, that in view of Eq. (\ref{eq:phi_qua})
the continuity of the field modes $\Phi_{\omega lm}(x)$ at the shell
implies the continuity of $G(\tilde{u},v;\varepsilon)$ there, therefore
\begin{equation}
G^{+}(\tilde{u}_{0},v_{0};\varepsilon)=G^{-}(\tilde{u}_{0},v_{0};\varepsilon)\,.\label{eq:G_equality}
\end{equation}
Also, since in our model $\left\langle \hat{\Phi}^{2}\right\rangle $
vanishes throughout $v<v_{0}$ as mentioned above, $\left\langle \hat{\Phi}^{2}\right\rangle _{\text{ren}}^{-}(x_{\text{sh}})$
vanishes and we can trivially replace $\left\langle \hat{\Phi}^{2}\right\rangle _{\text{ren}}^{+}(x_{\text{sh}})$
with $\left\langle \hat{\Phi}^{2}\right\rangle _{\text{ren}}^{+}(x_{\text{sh}})-\left\langle \hat{\Phi}^{2}\right\rangle _{\text{ren}}^{-}(x_{\text{sh}})$.
Using Eqs. (\ref{eq: theta_split-Sch}) and (\ref{eq: theta_split-Mink}),
and recalling Eq. (\ref{eq:G_equality}), we then get
\begin{equation}
\left\langle \hat{\Phi}^{2}\right\rangle _{\text{ren}}^{+}(x_{\text{sh}})=-\lim_{\varepsilon\to0}\left[G_{DS}^{+}(r_{0};\varepsilon)-G_{DS}^{-}(r_{0};\varepsilon)\right].\label{eq: G_difference}
\end{equation}
Note that when computing $\left\langle \hat{\Phi}^{2}\right\rangle _{\text{ren}}^{+}(x_{\text{sh}})$
directly from Eq. (\ref{eq: theta_split-Sch}), the most challenging
piece to evaluate is $G^{+}\left(\tilde{u}_{0},v_{0};\varepsilon\right)$,
as it requires solving the partial differential equations for the
individual field modes $\Phi_{\omega lm}(x)$, subsequently summing
and integrating over all mode contributions. Remarkably, this notoriously
demanding quantity is entirely eliminated in Eq. (\ref{eq: G_difference}),
leaving merely the simpler, well-known counterterms.

The explicit form of the $\theta$-splitting counterterm for $\left\langle \hat{\Phi}^{2}\right\rangle $
in Schwarzschild, as a function of the splitting parameter $\varepsilon$,
was computed in \citep{LeviOriTheta:2016}. Its general form is given
in Eq. (3.10) therein, in terms of the coefficients $a,c,d$ which
are specified in the equation preceding Eq. (4.2) therein. For Schwarzschild
with a general mass parameter $M$,
\[
G_{DS}(r;\varepsilon)=\hbar\left[\frac{1}{16\pi^{2}r^{2}}\sin^{-2}(\varepsilon/2)-\frac{M}{24\pi^{2}r^{3}}\right]\,.
\]
Substituting this expression in Eq. (\ref{eq: G_difference}) with
$M=M_{0}$ for $G_{DS}^{+}$ and $M=0$ for $G_{DS}^{-}$, and setting
$r=r_{0}$ in both, the shared $\varepsilon$-dependent term drops
out. Our result for $\left\langle \hat{\Phi}^{2}\right\rangle _{\text{ren}}$
at the shell's external side, in terms of $r_{0}$ (the $r$ value
on the shell), is therefore \footnote{Incidentally, when evaluating this result at $r_{0}=2M_{0}$, one
obtains $\hbar/192\pi^{2}M_{0}^{2}$, which is precisely the value
of $\left\langle \hat{\Phi}^{2}\right\rangle _{\text{ren}}$ at the
EH of a Schwarzschild BH in the Hartle-Hawking state \citep{Candelas:1980}.}
\begin{equation}
\left\langle \hat{\Phi}^{2}\right\rangle _{\text{ren}}^{+}(r_{0})=\frac{\hbar}{24\pi^{2}}\,\frac{M_{0}}{r_{0}^{3}}\,.\label{eq: result_Phi^2}
\end{equation}

\paragraph{Generalization to $\left\langle \hat{T}_{uu}\right\rangle _{\text{ren}}$.}

Since the shell is parameterized by $\tilde{u}$ (along with $\theta,\varphi$),
the continuity of $\Phi_{\omega lm}\left(x\right)$ across the shell
implies the continuity of $\Phi_{\omega lm,\tilde{u}}$ there, allowing
for a natural extension of the $\left\langle \hat{\Phi}^{2}\right\rangle _{\text{ren}}$
analysis to $\left\langle \hat{T}_{\tilde{u}\tilde{u}}\right\rangle _{\text{ren}}$
(and hence also to $\left\langle \hat{T}_{uu}\right\rangle _{\text{ren}}$).

The analog of Eq. (\ref{eq: PointSplit}) for the (trace-reversed
\footnote{The trace reverse of $T_{\alpha\beta}$ is defined as $T_{\alpha\beta}-g_{\alpha\beta}T_{\,\,\mu}^{\mu}/2$.})
RSET is given in Eq. (2.6) in \citep{LeviRSET:2017} (which applies
to \emph{any} splitting direction). When reduced to the $uu$ component
it reads  (recalling that $g_{uu}=0$ and hence $T_{uu}$ equals its
trace-reversed counterpart)
\begin{equation}
\left\langle \hat{T}_{uu}\right\rangle _{\text{ren}}(x)=\lim_{x'\to x}\left[G_{uu}(x,x')-\tilde{L}_{uu}(x,x')\right],\label{eq: PointSplit_Tuu}
\end{equation}
where $G_{uu}(x,x')\equiv\frac{1}{2}\left\langle \left\{ \hat{\Phi}_{,u}(x),\hat{\Phi}_{,u}(x')\right\} \right\rangle $,
$\left\{ ...\right\} $ denotes anti-commutation, and $\tilde{L}_{\mu\nu}(x,x')$
is the ``coordinate-based explicit counterterm'' obtained from Christensen's
counterterm $C_{\mu\nu}(x,x')$ as described in Eq. (2.5) in \citep{LeviRSET:2017}.

We shall again use $\theta$-splitting, placing $x$ at $\theta=0$
and $x'$ at $\theta=\varepsilon$. The $T_{uu}$ analog of Eq. (\ref{eq: theta_split})
now reads
\begin{equation}
\left\langle \hat{T}_{uu}\right\rangle _{\text{ren}}(u,v)=\lim_{\varepsilon\to0}\left[G_{uu}(u,v;\varepsilon)-\tilde{L}_{uu}(r;\varepsilon)\right]\,.\label{eq: theta_split_Tuu}
\end{equation}
We consider $\left\langle \hat{T}_{uu}\right\rangle _{\text{ren}}$
as the point $x_{\text{sh}}$ is approached from either the Minkowski
or Schwarzschild side (i.e., $x\to x_{\text{sh}}^{\pm}$). We obtain
the $\left\langle \hat{T}_{uu}\right\rangle _{\text{ren}}$ analog
of Eqs. (\ref{eq: theta_split-Sch}) and (\ref{eq: theta_split-Mink}),
which we collectively write as
\begin{equation}
\left\langle \hat{T}_{uu}\right\rangle _{\text{ren}}^{\pm}(x_{\text{sh}})=\lim_{\varepsilon\to0}\left[G_{uu}^{\pm}(\tilde{u}_{0},v_{0};\varepsilon)-\tilde{L}_{uu}^{\pm}(r_{0};\varepsilon)\right].\label{eq: Tuu_sch,mink,u}
\end{equation}

We would again like to subtract the ``$-$'' version of $\left\langle \hat{T}_{uu}\right\rangle _{\text{ren}}$
from its ``$+$'' version, to eliminate the $G_{uu}$ terms altogether.
Recall, however, that it is $\hat{\Phi}_{,\tilde{u}}$ that is continuous
across the shell, not $\hat{\Phi}_{,u}$. Therefore, $G_{uu}$ is
not continuous at the shell, whereas $G_{\tilde{u}\tilde{u}}$ is,
where $G_{\tilde{u}\tilde{u}}(x,x')\equiv\frac{1}{2}\left\langle \left\{ \hat{\Phi}_{,\tilde{u}}(x),\hat{\Phi}_{,\tilde{u}}(x')\right\} \right\rangle $.
This motivates transforming Eq. (\ref{eq: Tuu_sch,mink,u}) from $u$
to $\tilde{u}$, via multiplication by $(du/d\tilde{u})^{2}$:
\begin{equation}
\left\langle \hat{T}_{\tilde{u}\tilde{u}}\right\rangle _{\text{ren}}^{\pm}(x_{\text{sh}})=\lim_{\varepsilon\to0}\left[G_{\tilde{u}\tilde{u}}^{\pm}(\tilde{u}_{0},v_{0};\varepsilon)-\tilde{L}_{\tilde{u}\tilde{u}}^{\pm}(r_{0};\varepsilon)\right],\label{eq: Tuu_sch,mink,utilde}
\end{equation}
where
\[
\tilde{L}_{\tilde{u}\tilde{u}}^{\pm}(r_{0};\varepsilon)\equiv\left(\frac{du}{d\tilde{u}}\right)^{2}\tilde{L}_{uu}^{\pm}(r_{0};\varepsilon)\,.
\]
Recalling Eq. (\ref{eq: du_tilde}) (and $du/d\tilde{u}=1$ at $v>v_{0}$),
we get
\begin{equation}
\tilde{L}_{\tilde{u}\tilde{u}}^{+}(r_{0};\varepsilon)=\tilde{L}_{uu}^{+}(r_{0};\varepsilon)\,,\,\,\,\,\tilde{L}_{\tilde{u}\tilde{u}}^{-}(r_{0};\varepsilon)=f_{0}^{2}\tilde{L}_{uu}^{-}\label{eq: L_utilde}
\end{equation}
where $f_{0}\equiv1-2M_{0}/r_{0}$.

From this point, the analysis proceeds in full analogy with the above
treatment of $\left\langle \hat{\Phi}^{2}\right\rangle _{\text{ren}}$.
First, recalling that $\left\langle \hat{T}_{\tilde{u}\tilde{u}}\right\rangle _{\text{ren}}$
(like $\left\langle \hat{T}_{uu}\right\rangle _{\text{ren}}$) vanishes
in the Minkowski region,
\begin{equation}
\left\langle \hat{T}_{\tilde{u}\tilde{u}}\right\rangle _{\text{ren}}^{+}(x_{\text{sh}})=\left\langle \hat{T}_{\tilde{u}\tilde{u}}\right\rangle _{\text{ren}}^{+}(x_{\text{sh}})-\left\langle \hat{T}_{\tilde{u}\tilde{u}}\right\rangle _{\text{ren}}^{-}(x_{\text{sh}})\,.\label{eq: Tuu_difference}
\end{equation}
Next, the continuity of $G_{\tilde{u}\tilde{u}}(x,x')$ across the
shell implies that $G_{\tilde{u}\tilde{u}}^{+}(\tilde{u}_{0},v_{0};\varepsilon)=G_{\tilde{u}\tilde{u}}^{-}(\tilde{u}_{0},v_{0};\varepsilon)$,
allowing us to rewrite Eq. (\ref{eq: Tuu_sch,mink,utilde}) as
\[
\left\langle \hat{T}_{\tilde{u}\tilde{u}}\right\rangle _{\text{ren}}^{\pm}(x_{\text{sh}})=\lim_{\varepsilon\to0}\left[G_{\tilde{u}\tilde{u}}(\tilde{u}_{0},v_{0};\varepsilon)-\tilde{L}_{\tilde{u}\tilde{u}}^{\pm}(r_{0};\varepsilon)\right]\,,
\]
and substituting it in Eq. (\ref{eq: Tuu_difference}) we obtain
\[
\left\langle \hat{T}_{\tilde{u}\tilde{u}}\right\rangle _{\text{ren}}^{+}(x_{\text{sh}})=-\lim_{\varepsilon\to0}\left[\tilde{L}_{\tilde{u}\tilde{u}}^{+}(r_{0};\varepsilon)-\tilde{L}_{\tilde{u}\tilde{u}}^{-}(r_{0};\varepsilon)\right]\,.
\]
At this stage, it becomes convenient to re-express this equation in
terms of the Eddington $u$ coordinate instead of $\tilde{u}$, using
Eq. (\ref{eq: L_utilde}) and recalling that $\left\langle \hat{T}_{\tilde{u}\tilde{u}}\right\rangle _{\text{ren}}^{+}=\left\langle \hat{T}_{uu}\right\rangle _{\text{ren}}^{+}$:
\begin{equation}
\left\langle \hat{T}_{uu}\right\rangle _{\text{ren}}^{+}(x_{\text{sh}})=-\lim_{\varepsilon\to0}\left[\tilde{L}_{uu}^{+}(r_{0};\varepsilon)-f_{0}^{2}\tilde{L}_{uu}^{-}(r_{0};\varepsilon)\right]\,.\label{eq: Luu_difference}
\end{equation}
Both $\tilde{L}_{uu}^{+}(r_{0};\varepsilon)$ and $\tilde{L}_{uu}^{-}(r_{0};\varepsilon)$
are obtained from the basic $\theta$-splitting counterterm $\tilde{L}_{uu}(r;\varepsilon)$
for the Schwarzschild metric of general mass parameter $M$, by substituting
the appropriate mass parameter ($M_{0}$ or $0$) and setting $r=r_{0}$.
The expression in square brackets in Eq. (\ref{eq: Luu_difference})
then becomes
\begin{equation}
\left(\tilde{L}_{uu}(r_{0};\varepsilon)|_{M=M_{0}}\right)-f_{0}^{2}\left(\tilde{L}_{uu}(r_{0};\varepsilon)|_{M=0}\right)\,.\label{eq: Luu_difference-1}
\end{equation}
The mentioned basic Schwarzschild's counterterm $\tilde{L}_{uu}(r;\varepsilon)$
was presented in Sec. I of the Supplemental Material of \citep{FluxesRN:2020}
(setting $Q=0$). It reads 
\begin{equation}
\tilde{L}_{uu}(r;\varepsilon)=\hbar\left[\frac{f^{2}}{64\pi^{2}r^{4}}\sin^{-2}(\varepsilon/2)+e_{uu}\right]\,,\label{eq: L_Schwarz}
\end{equation}
where $e_{uu}=-3Mf{}^{2}/160\pi^{2}r^{5}$. Substituting this expression
into Eq. (\ref{eq: Luu_difference-1}), the two terms $\propto\sin^{-2}(\varepsilon/2)$
cancel out, leaving only the $\propto e_{uu}$ contribution \textendash{}
which is independent of $\varepsilon$ (and vanishes in the $M=0$
case). This yields our final result for the outflux density at the
shell's external side (at a point where $r=r_{0}$):
\begin{equation}
\left\langle \hat{T}_{uu}\right\rangle _{\text{ren}}^{+}\left(r_{0}\right)=\hbar\frac{3M_{0}}{160\pi^{2}r_{0}^{5}}\left(1-\frac{2M_{0}}{r_{0}}\right)^{2}\,.\label{eq: result_Tuu}
\end{equation}

\paragraph{Discussion.}

The result (\ref{eq: result_Tuu}) reveals that as the shell approaches
the EH, the outflux density $\left\langle \hat{T}_{uu}\right\rangle _{\text{ren}}^{+}$
emerging from the shell vanishes quadratically with the remaining
distance $r_{0}-2M_{0}$. This vanishing rate ensures the regularity
of the Kruskal-based outflux density $\left\langle \hat{T}_{UU}\right\rangle _{\text{ren}}^{+}$
at the EH. \foreignlanguage{american}{Since the RSET magnitude is
typically of order $\hbar/M^{2}\lll1$, }this regularity actually
\foreignlanguage{american}{implies that semiclassical backreaction
\emph{cannot} prevent (or significantly affect) the  horizon formation.}

Our result clarifies the origin of the Hawking outflux (at least in
the collapsing null shell case). Consider the evolution of $F=4\pi r^{2}\left\langle \hat{T}_{uu}\right\rangle _{\text{ren}}$
along an outgoing null ray $u=\text{const}\equiv u_{0}\gg M_{0}$.
This geodesic leaves the shell at $r=r_{0}\approx2M_{0}$ and propagates
toward FNI. Equation (\ref{eq: result_Tuu}) implies that the outflux
emerging from the shell, $F^{+}\left(r_{0}\right)\equiv4\pi r_{0}^{2}\left\langle \hat{T}_{uu}\right\rangle _{\text{ren}}^{+}\left(r_{0}\right)$,
is negligibly small for such a geodesic; whereas when the geodesic
approaches the weak-field region ($r\gg M_{0}$), $F$ has already
reached its asymptotic value $F_{H}$. That is, the evolution of $F$
along the outgoing null ray from $F^{+}\left(u_{0}\right)\approx0$
(at the shell) to $F_{H}$ (at FNI) occurs, gradually, as the ray
traverses the strong-field region. This designates scenario (\emph{i}),
presented in the Introduction, as the correct physical description.

Multiplying Eq. (\ref{eq: result_Tuu}) by $4\pi r_{0}^{2}$, one
finds that at $r_{0}^{\text{max}}=(10/3)M$ the mentioned outflux
$F^{+}\left(r_{0}\right)$ reaches a maximum value of $\left(81/250000\pi\right)\hbar M^{-2}\approx1.031\times10^{-4}\hbar M^{-2}$.
Interestingly, this value exceeds the known \citep{Elster:1983} Hawking
outflux in Schwarzschild, $F_{H}\approx7.44\times10^{-5}\hbar M^{-2}$.
This suggests that perhaps\textcolor{red}{{} }the outflux $F$ evolves
\emph{nonmonotonically} (with $r$) along the outgoing null geodesic
emanating from the shell at $r_{0}^{\text{max}}$. This should not
be too surprising, as such nonmonotonic behavior of the outflux is
actually observed already in the Unruh state in Schwarzschild (see
\citep{Sup}).

As noted in footnote {[}41\ref{fn:transition}{]}, our computational
method inherently involves an interchange of two limits, which we
have not attempted to rigorously justify. A deeper mathematical investigation
of this issue of interchangeability would be desired, although such
an analysis may be rather challenging. Generally speaking, the computation
of renormalized quantities in BH backgrounds may typically involve
several limits, including: (i) the coincidence limit $x'\to x$, (ii)
an infinite mode sum (if mode decomposition is used), and (iii) in
some cases, an additional limit of approaching a desired destination
point $x_{\text{dest}}$ (e.g., when a direct computation at $x_{\text{dest}}$
is problematic or challenging). Basically, the required order of limits
is as follows: first, performing the mode sum (if relevant); second,
taking the coincidence limit (after counterterm subtraction); and
finally, approaching the destination point (when applicable). However,
in various concrete computational schemes, some of these limits are
often interchanged. For instance, in the ``Pragmatic Mode-sum Regularization''
method \citep{LeviOriT:2015,LeviOriRSET:2016}, the mode sum and the
coincidence limit are effectively interchanged. Nevertheless, results
obtained through this method have been separately confirmed in multiple
cases \citep{LeviOriT:2015,LeviOriRSET:2016,LeviRSET:2017,LeviEilonOriMeent:2017,SchLanir:2018,KerrIH:2022,t-splitKerr:2024}.

In the present analysis, the coincidence and the $x\to x_{\text{dest}}$
limits were implicitly interchanged. This situation somewhat resembles
the flux computation at a Kerr BH's inner horizon in Ref. \citep{KerrIH:2022},
where the mode sum and the limit $r\to r_{-}$ were effectively interchanged:
while $\left\langle T_{vv}\right\rangle _{\text{ren}}^{-}$ was defined
there as $\lim_{r\to r_{-}}\left\langle T_{vv}\right\rangle _{\text{ren}}\left(r\right)$
(mainly because Unruh's field modes are literally ill-defined at $r=r_{-}$),
in practice this $r\to r_{-}$ limit was applied to the individual
mode contributions \emph{before} summation and integration. The validity
of the results in that case was nevertheless independently confirmed
by applying $t$-splitting at a set of points reaching very close
to $r_{-}$ \citep{t-splitKerr:2024}. Similarly in our case, it would
be highly beneficial to conduct an independent computation of the
outflux emerging from the shell (e.g. by using the methods developed
in Refs. \citep{AndersonSiahmazginClarkFabbri:2020,SiahmazgiAndersonClarkFabbri:2021}),
to provide a robust cross-check of our result (\ref{eq: result_Tuu}).

\selectlanguage{american}%
In the analogous 2D case the RSET is known explicitly \citep{Hiscock:1981}\citep{FabbriNavarroSalas:Book},
and one can easily see that $\left\langle T_{\tilde{u}\tilde{u}}\right\rangle $
actually vanishes at both sides of the collapsing null shell \textendash{}
in contrast with the discontinuity of $\left\langle T_{\tilde{u}\tilde{u}}\right\rangle $
at the shell in 4D. This sharp distinction may be rooted in two obvious
differences between the two cases. First and most importantly, in
the 2D case $\left\langle T_{\alpha}^{\alpha}\right\rangle $ is known
(from the trace anomaly \citep{Davies:1977}) to be proportional to
the background's Ricci scalar \textendash{} which is bounded at the
shell. From this boundedness one can easily show (using energy-momentum
conservation) that $\left\langle T_{\tilde{u}\tilde{u}}\right\rangle $
\emph{must be continuous at the null shell}. Such a constraint does
not apply in 4D, primarily because $\left\langle T_{\alpha\beta}\right\rangle $
now involves one additional degree of freedom (related to the angular
RSET components). Second, recall that in 4D our minimally-coupled
field $\Phi(x)$ is non-conformal, hence $\left\langle T_{\alpha}^{\alpha}\right\rangle $
now includes an extra term $\propto\boxempty\langle\Phi^{2}\rangle$
\citep{Group:2018} \textendash{} which in fact involves $\delta(v-v_{0})$.
In view of this delta function in $\left\langle T_{\alpha}^{\alpha}\right\rangle $
in 4D, a simple analysis (involving energy-momentum conservation)
leads to the anticipation for a step function in $\left\langle T_{\tilde{u}\tilde{u}}\right\rangle $
at the shell \textendash{} which was indeed found here explicitly.

\selectlanguage{english}%
It may be interesting \textendash{} and physically important \textendash{}
to extend the analysis to the more realistic case of an electromagnetic
field.
\begin{acknowledgments}
We are grateful to Paul Anderson, Jochen Zahn, Stefan Hollands, Marc
Casals, Adrian Ottewill and Maria Alberti for interesting discussions
and helpful feedback. We are especially thankful to Adam Levi for
providing the digital version of the Unruh-state RSET data of Ref.
\citep{LeviRSET:2017} used for the preparation of Fig. {[}1{]} in
the Supplemental Material.

N.Z. acknowledges the generous support of Fulbright Israel.
\end{acknowledgments}

\end{document}